\documentclass[reprint,amsmath, amssymb, aps, longbibliography, prl, article]{revtex4-2}
\usepackage[usenames, dvipsnames]{color}
\usepackage{tabto,lipsum} % add margin symbols
\usepackage{floatrow} % add side captions

\usepackage{hyperref}
\hypersetup{
	colorlinks=true,
	linkcolor=blue,
	citecolor=blue,
	filecolor=blue,
	urlcolor=blue,
}

\usepackage{graphicx} 
\usepackage{svg,svg-extract}

\usepackage{amsmath} 
\usepackage{amssymb}
\usepackage{braket}
\usepackage{upgreek}

\usepackage[normalem]{ulem}
\usepackage[toc,page]{appendix}

\usepackage{xcolor}

\begin{document}
	
	\title{Super-resolution optical trapping of multiple cold atoms}
	
	\author{Kelvin Lim$^{1}$}
	\author{Vincent Mancois$^{1,2}$}%
	\altaffiliation[corresponding author: ]{vincent.mancois@ntu.edu.sg}
	\author{Haijun Wu$^{3}$}
	\author{Yijie Shen$^{1,4}$}
	\author{David Wilkowski$^{1,2,5}$}
	\affiliation{ 
		$^1$ Centre for Disruptive Photonic Technologies, SPMS, Nanyang Technological University, 637371, Singapore.\\
		$^2$ MajuLab, International Joint Research Unit IRL 3654, CNRS, Université Côte d’Azur, Sorbonne Université, National University of Singapore, Nanyang Technological University, Singapore.\\
		$^3$ Wang Da-Heng Center, Heilongjiang Key Laboratory of Quantum Control, Harbin University of Science and Technology, Harbin 150080, China\\
		$^4$ School of Electrical and Electronic Engineering, Nanyang Technological University, Singapore\\
		$^5$ Centre for Quantum Technologies, National University of Singapore, 117543, Singapore.
	}%

	\begin{abstract}
		Arrays of optical tweezers form the backbone of neutral atoms analog and digital quantum processors. However, the inter-trap distance remains generally much larger than the size of the tweezers to avoid interference-induced trap distortions, limiting the trap density. Here, we report single-atom trapping in four super-resolved tweezers, meaning with a separation below the Sparrow diffraction limit. The optical pattern is generated using superoscillatory phenomenon leading to subwavelength traps with full control of the trap relative phases. We investigate two sets of relative phases that impede or allow the hopping and the reshuffling of atoms. We envision that superoscillatory light structuring will bridge the gap between large-distance traps generated by tweezer arrays and short-distance traps formed with optical lattices.
	\end{abstract}
	\date{\today}
	
	\maketitle
	
	The trapping of neutral atoms in laser beams via dipolar force is a ubiquitous technique used for quantum computing and simulation. We can roughly divide optical dipole traps (ODTs) into two types: optical lattices \cite{Hemmerich1993} and optical tweezers \cite{Ashkin1986}. For an optical lattice, the periodic inter-trap separation can reach its minimal value of $\lambda/2$, where $\lambda$ is the wavelength of the laser field. The resilience of some optical-lattice configurations in trap depth and trap geometry to phase fluctuations makes them extremely valuable in the context of quantum simulation \cite{gross2017quantum}. Tweezer arrays, on the other hand, give access to single-atom positioning at the cost of larger trap separations as ODTs are generated by an imaging system with numerical aperture (NA) that is smaller than one in vacuum. In most applications, the different sites of the tweezer arrays are far enough so that their phase remains unconstrained and that phase retrieval can be performed to homogenize their intensity distribution, resulting in thousands of ODTs \cite{barredo2016,endres2016,manetsch2024} with promising applications for neutral atoms digital \cite{bluvstein_logical_2024} and analog \cite{scholl2021quantum,ebadi_quantum_2021} quantum computing. Despite the strong interest in the current tweezer arrays for quantum technologies, salient perspectives are expected by further reducing the inter-trap separation and, consequently, the ODT size, or vice versa. For example, more ODTs can be packed in the field of view of the imaging system, and a larger Rydberg interaction range is expected. If the inter-trap separation becomes comparable to or smaller than the wavelength, a cooperative emission of light emerges with growing interest for quantum information \cite{PhysRevLett.122.203605,RevModPhys.82.1041}, and the creation of light-atom interfaces \cite{PhysRevA.78.053816, PhysRevLett.111.147401, rui2020subradiant, PhysRevLett.118.113601}. Tweezer arrays are also important tools for quantum simulations to obtain highly correlated low-entropy states \cite{Norcia2018,Spar2022}. It is also long known that reducing the inter-trap separation \cite{Yi2008} in turn increases the tunneling energy, which speeds up quantum simulations. More recently, quantum many-body scars \cite{Bernien2017, Turner2018, Serbyn2021, Bluvstein2021, Zhang2022, Su2023} have rekindled the interest for few-body quantum simulations in one dimension (1D) with periodic or open boundary conditions \cite{Moudgalya2022}, calling for improvements in tweezer tunability rather than scalability. To address the challenging problem of short inter-trap distance, one must first reduce the ODT size below the Abbe limit. This has been demonstrated using the super-oscillatory (SO) phenomenon \cite{Rogers2018}, where the trapping of a single atom in sub-wavelength tweezers has been reported \cite{rivy2023single}. In addition, as the inter-trap separation becomes comparable to the ODT size, interference effects modify the optical intensity pattern, requiring control of the trap relative phase, for example, using new phase retrieval algorithms \cite{nishimura2024}.\\
	\indent In this Letter, we report on a cluster of four super-resolved optical tweezers fulfilling the above-mentioned requirements, namely a super-resolved regime and trap size smaller than the Abbe limit. The pattern is generated with an SO field combined with controlled relative phases of the tweezers. We propose two configurations of the relative phases of the tweezers that impede or allow the hopping of atoms between tweezers. Initially, the atoms are cooled in a magneto-optical trap (MOT) and loaded into standard Airy tweezers before being continuously transferred into the super-resolved SO tweezers. Finally, the atoms are transferred back into the Airy tweezers for analysis of the survival probability and the inter-trap reshuffling mechanisms, dominated here by the thermal energy of the atoms. The atom reshuffling obeys a simple combinatorial model that describes the observed atom dynamics. We further demonstrate the progressive loss of memory of the single atom loading position by lowering the inter-trap potential barrier.
	
	\begin{figure}[t]
		%%%%%% Figure 1 %%%%%%%%%%%%%%%%%%%%
		\includegraphics[width=1\textwidth]{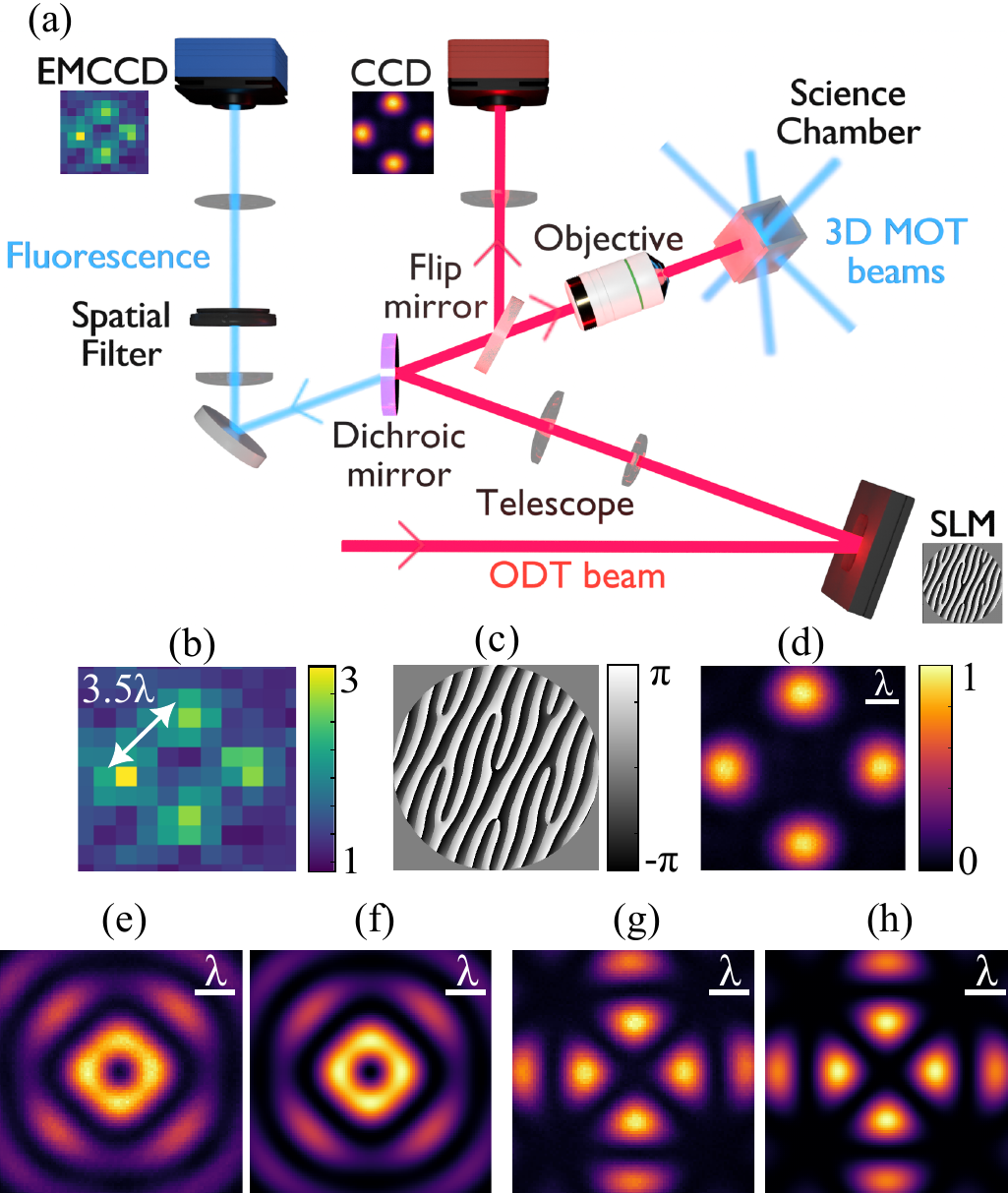}
		\caption{(a) Sketch of the experiment that shows the 3D MOT beams, the fluorescence path, and the ODT beam path at $1064\,$nm. The imaging system is diffraction-limited at $1064\,$nm, with NA = 0.39. The insets illustrate image samples obtained on CCD cameras and a phase pattern loaded into the SLM. They are presented in (b)-(d) with greater resolutions. (b) Fluorescence image of four atoms in the Airy tweezers with $3.5\lambda$ separation. (c) The Airy hologram loaded into the SLM. For clarity purposes, the period of the blazed grating is represented with a multiplication factor of 10. (d) Airy pattern corresponding to the hologram depicted in (c). (e) and (f) [(g) and (h)] Measured and simulated SO light pattern for $2\pi$ ($4\pi$) winding of the trap relative phase. The scale in (d)-(h) corresponds to the object plane located in the science chamber.}
		\label{fig:patterns}
	\end{figure}
	The experiment starts by loading a MOT of Cesium (Cs) atoms in the presence of ODTs consisting of four Airy tweezers. Figure \hyperref[fig:patterns]{1(a)} sketches the key elements of the experiment: the ODT laser at a wavelength of $\lambda=1064\,$nm, the MOT beams red detuned with respect to the Cs $D_2$ line at $852\,$nm, the fluorescence path, and the imaging system. The trapped atoms are imaged using an EMCCD camera with a magnification factor of $14.7\times$. Figure \hyperref[fig:patterns]{1(b)} shows a single shot fluorescence image of four atoms in the Airy tweezers with an inter-trap separation of $3.5\lambda$. This separation is an optimum compromise between a small separation that facilitates the transfer to the super-resolved tweezer array and a large separation for well-resolved fluorescence images, (see Supplemental Material \cite{SM}). A spatial light modulator (SLM), located in a Fourier plane of the MOT, creates the multi-trap profiles by performing amplitude and phase shaping of the ODT beam transverse profile, using a single phase hologram \cite{Guzman2017}. It consists of a blazed grating and four superimposed linear phase ramps. The slope of each ramp imposes the angular deviation of a beam from the optical axis, whereas the initial value of the ramp fixes the phase of the ODT. Figure \hyperref[fig:patterns]{1(c)} shows the hologram that generated the four Airy tweezers. The latter are imaged using an electron multiplying CCD (EMCCD) camera with a magnification factor of $49\times$, [see  Fig. \hyperref[fig:patterns]{1(d)}]. For the SO profiles, the depth of the phase mask is modulated by an amplitude mask consisting of a ring-shaped aperture \cite{SM}. Our SO hologram is a simple extension of an annular beam \cite{Duocastella2012} to a tweezer array.\\
	\indent We load atoms into Airy tweezers before transferring them into the SO tweezers for unambiguous trapping positions at the central spot. In fact, SO tweezers, in addition to the central spot, possess outer rings that can potentially trap atoms \cite{rivy2023single}. Once the tweezers are loaded, the SLM hologram is changed progressively to transfer the atoms into the SO tweezers and ultimately into the final super-resolved pattern. We verified that using a single intermediate hologram is sufficient to obtain a transfer probability close to unity, taking advantage of the slow settling time of the liquid crystal of the SLM \cite{SM}.\\
	\indent As the tweezers get closer, the influence of their relative phase on the intensity pattern becomes significant. We show in Figs. \hyperref[fig:patterns]{1(e)} and \hyperref[fig:patterns]{1(g)} the measured final super-resolved configurations studied here, corresponding to a $2\pi$-phase winding $\phi_k= k\pi/2$ and $4\pi$-phase winding $\phi_k=k\pi$. $k$ is an integer labeling the trap, for example, in the clockwise order. We note that the two phase-winding sets are compatible with the desired four-fold rotation symmetry of the tweezer array pattern. We exclude the $\phi_k=0$ configuration as the resulting constructive interference leads to the unwanted merging of the tweezers. In contrast, the $4\pi$-winding case, corresponding to destructive interference between neighboring tweezers, exhibits high-contrast traps. Finally, the $2\pi$-winding solution lets the tweezers move closer to each other with a finite potential barrier of $\sim 15\%$ of the total barrier height, [see Fig. \hyperref[fig:patterns]{1(e)}]. We measured a relative standard deviation of $\sim 3\%$ in the peak intensity for both windings. We note that the experimental pattern is in excellent agreement with the theoretical one, depicted in Figs. \hyperref[fig:patterns]{1(f)} and \hyperref[fig:patterns]{1(h)}, computed in the paraxial and scalar approximations using a Collins integral \cite{collins1970}.
	
	\begin{figure}[t]
		%%%%%% Figure 2 %%%%%%%%%%%%%%%%%%%%
		\includegraphics[width=0.9\textwidth]{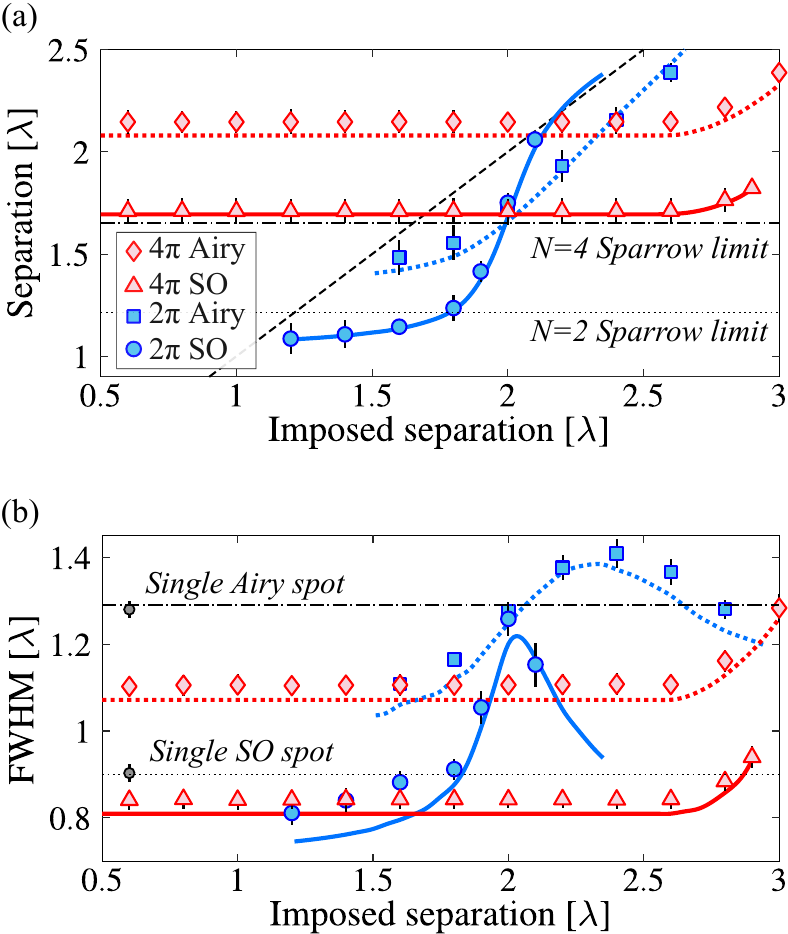}
		\caption{(a) Neighboring trap separation, and (b) tweezer radial FWHM size. The $2\pi$- ($4\pi$-) winding cases are in blue (red), whereas the diamond and square (triangle and circle) symbols represent measurements of the Airy (SO) tweezers. The colored solid and dotted curves are simulations using Collins integrals without free parameters. The black dashed line in (a) corresponds to ideal point-like tweezers where the imposed separation is equal to the measured separation. The simulated Sparrow limits for two tweezers (dotted line) and four tweezers (dash-dotted line) are shown in (a). The measured (black dot) and expected (black lines) FWHM sizes of a single Airy and SO tweezers are shown in (b).}
		\label{fig:sep_fwhm}
	\end{figure}
	
	To characterize in more detail the phase-winding configurations, we plot the tweezer separation in Fig. \hyperref[fig:sep_fwhm]{2(a)} as a function of the imposed separation. The imposed separation is used as a parameter to define the hologram phase ramps and would correspond to the expected separation for point-like tweezers (dashed black line). 
	We indicate the two-spot (respectively four-spot) Sparrow limit given by $0.47\lambda/$NA (respectively $0.64\lambda/$NA), corresponding to the separation where two (respectively four) incoherent tweezers merge. The $2\pi$-winding Airy and SO profiles (blue symbols) lie below the four-spot Sparrow limit, indicating a super-resolved regime. In contrast, the $4\pi$-winding cases (red symbols) remain above the four-spot Sparrow limit, illustrating the key role of trap interferences. We note that the separation of the SO tweezers is lower than that of the Airy ones for both windings.
	
	\begin{figure}[t!]
		%%%%%% Figure 3 %%%%%%%%%%%%%%%%%%%%
		\includegraphics[width=1.0\textwidth]{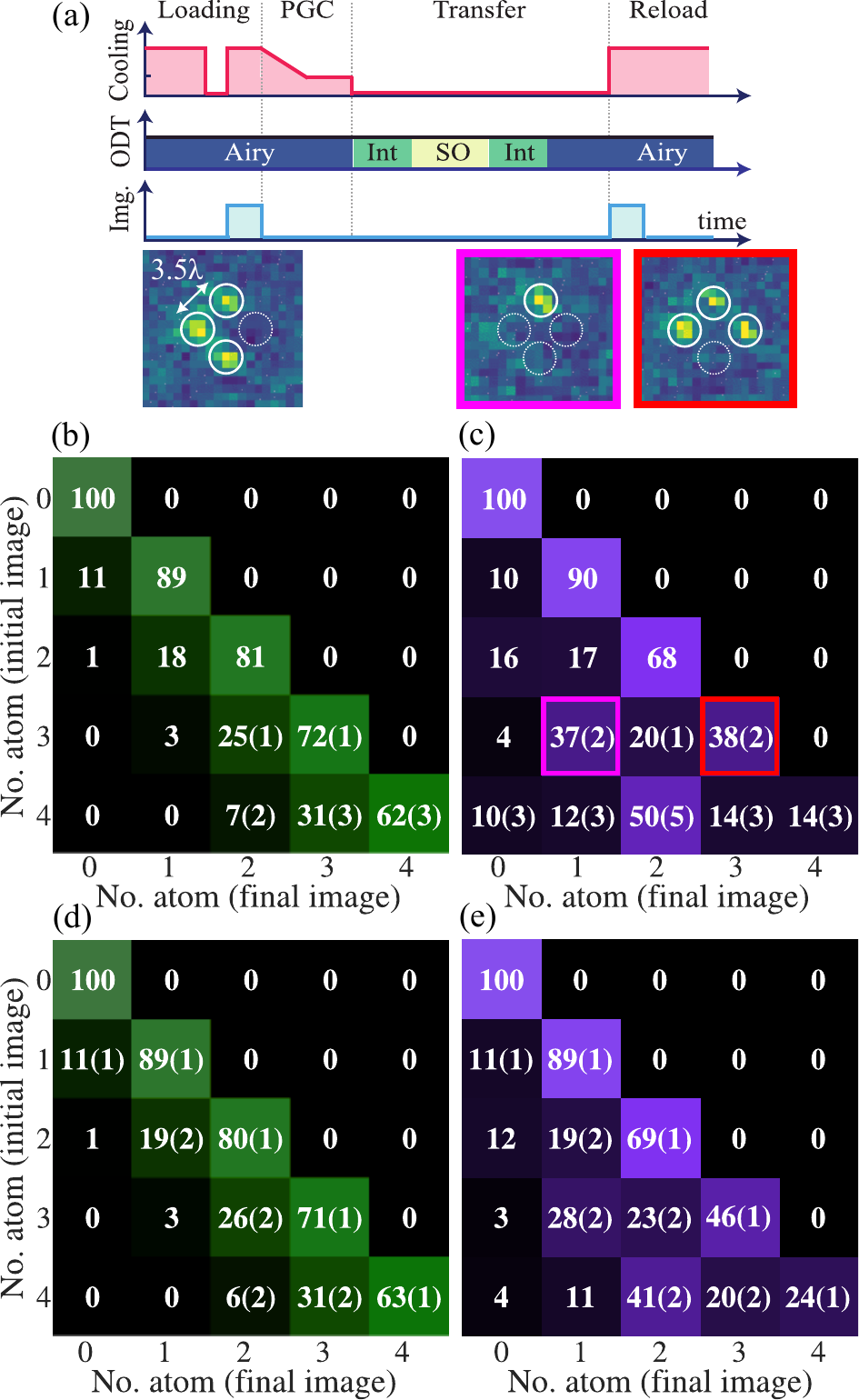}
		\caption{(a) Schematic of the time sequences, comprising four phases: a loading and initial imaging, a polarization gradient cooling (PGC), an Airy-SO-Airy transfer, and a final imaging (see Supplemental Material \cite{SM} for more details). A four-atom probability matrix after an Airy-SO-Airy round trip for the case of (b) $4\pi$-winding and (c) the $2\pi$-winding.
		The theory predictions are shown in (d) and (e). The matrix elements $M_{i,j}$ give the probability to measure $i$ ($j$) atoms in the initial (final) image. The index $i$ ($j$) corresponds to the row (column). The $M_{i,j}$ values are given in percentage normalized for each initial atom number, regardless of the arrangement of the atoms in the traps. The image panels illustrate a case where three atoms are loaded (left image). After transfer, this case might contribute to $M_{3,1}$ (right image highlighted in fuchsia) due for example to a two-body loss. A contribution to the diagonal term $M_{3,3}$ is shown in the right image highlighted in red. Each experimental matrix is computed using $10^4$ runs. The standard deviations above 1$\%$ are indicated and large values mainly reflect the low occurrence of a realization. The uncertainty for the theory in (d) and (e) comes from the statistical incertitude on the temperature and atom lifetime.}
		\label{fig:matrices}
	\end{figure}
	
	The analysis of phase-winding configurations is extended to the radial full width at half maximum (FWHM) of the ODTs, and we show the main results in Fig. \hyperref[fig:sep_fwhm]{2(b)}. Here, the dashed-dotted line is $\lambda/2\textrm{NA}$, corresponding to the Abbe limit or the FWHM of an Airy spot, whereas the dotted line is the predicted FWHM size of a single SO spot that is $\sim 30\%$ smaller than that of the Abbe limit. The black circles are the measured Airy and SO single spot in excellent agreement with the predicted values. Interestingly, at short imposed separations ($<1.8\lambda$), the four-trap Airy configuration lies below the single spot Airy, and so does the four-trap SO configuration. Finally, we stress that for SO-traps, the FWHM size is sub-wavelength. The curves in Fig. \hyperref[fig:sep_fwhm]{2} correspond to the simulations using the Collins integral and are in good agreement with the measurements. Nevertheless, the predicted FWHMs are generally below the experimental data. It might indicate residual uncompensated aberrations in the imaging system that do not seem to affect the single ODTs \cite{SM}.\\
	\indent In the remainder of this Letter, we analyze the behavior of loaded atoms in the SO tweezers. A schematic of the experimental sequence is shown in Fig. \hyperref[fig:matrices]{3(a)}. After loading, the initial atom distribution is measured via fluorescence imaging [see Fig. \hyperref[fig:patterns]{1(b)}]. We then dynamically change the SLM hologram from Airy to SO, then we hold the atoms for 20 ms in the SO tweezers and finally transfer back to Airy, before a final fluorescence imaging to assess atom losses and reshuffling. In the SO tweezers, the trap depth is $74.8(7)\,\upmu$K for the $2\pi$-winding case and $125(1)\,\upmu$K for the $4\pi$-winding case. In the $2\pi$-winding case, the traps are linked by a small potential barrier of $9.0(5)$ $\upmu$K, while in the $4\pi$-winding case the spots remain well separated. The lifetime of a single atom in the Airy tweezers is measured to be $\tau=1.8(3)\,$s, limited by the residual background pressure. The time duration between the initial and final images is $t_i=210\,$ms. Thus, the survival probability of an atom after a round trip is upper bounded by $p_b=\exp{(-t_i/\tau)}\simeq 89(1)\%$.
	
	\begin{figure}[t!]
		%%%%%% Figure 4 %%%%%%%%%%%%%%%%%%%%
		\includegraphics[width=1\textwidth]{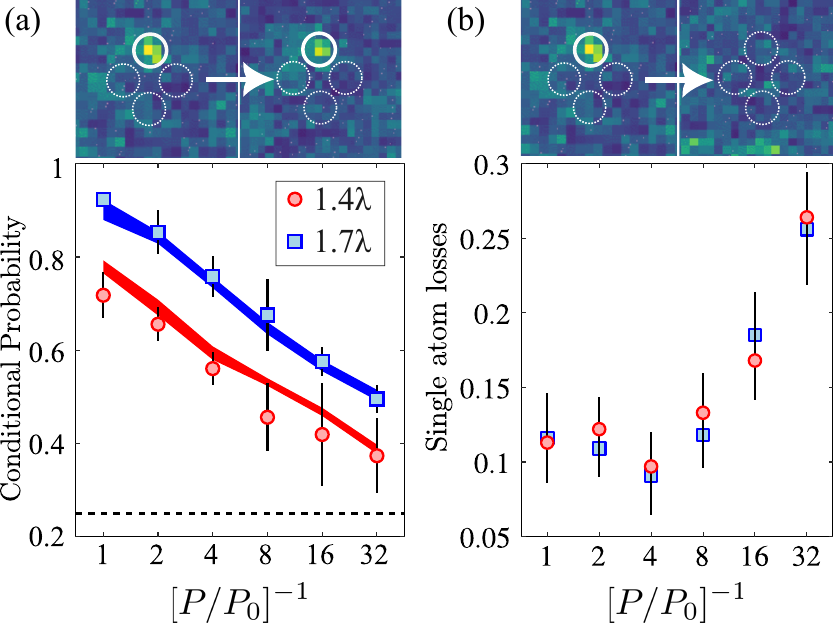}
		\caption{(a) Conditional probability for an atom to remain in the same trap after an Airy-SO-Airy round trip as a function of an adiabatic reduction in tweezer power during $60\,$ms in the SO tweezer phase. $P_{0}$ (respectively $P$) is the SO tweezer power before (after) reduction. An example of a positive count is illustrated on the image panels above. The dashed black line indicates the conditional probability of a complete reshuffling. The conditional probability is derived from simulations and shown as shaded areas for a 1.4$\lambda$ (red) and a 1.7$\lambda$ (blue) trap separation. The solid area width represents a standard deviation error that comes from the initial temperature uncertainty. (b) one-body survival probability as a function of the reduction in tweezer power. An atom loss is illustrated in the image panels above. In (a) and (b) the red circles (blue squares) correspond to a trap separation of 1.4$\lambda$ (1.7$\lambda$). Each data point corresponds to an average over $\sim180$ occurrences, and the error bars correspond to the standard deviations.}
		\label{fig:decomp}
	\end{figure}

	In Figs. \hyperref[fig:matrices]{3(b)} and \hyperref[fig:matrices]{3(c)}, we show the survival probability matrix after a round trip for the $4\pi$- and $2\pi$-winding cases, respectively. The matrix elements $M_{i,j}$ give the probability to measure $i$ ($j$) atoms in the initial (final) image. The values are given in percentage, normalized for each $i$ value. We note that $M_{i<j}=0$. This is expected as the tweezers are loaded before the first image and it also signifies that the false count is below the $1\,\%$ level \cite{SM}. $M_{i> j}\neq 0$ indicates that losses occur.
	One-body losses come from residual background pressure or can be due to the spilling atoms, as the trap depth varies during the transfer. We also expect a two-body loss mechanism coming from light-induced inelastic losses \cite{schlosser2001sub} during the final fluorescence imaging. This mechanism removes atoms by pair, the atom number becomes zero (respectively one) if the trap contains an even (respectively odd) number of atoms \cite{sherson2010single}. If such a two-body loss mechanism is at play, it indicates the hopping and reshuffling of the atoms in the traps after a round trip. For the $4\pi$-winding case, the matrix elements are well-captured by our model [see Fig. \hyperref[fig:matrices]{3(d)}] considering only one-body loss, namely  $M_{i,j\leq i}(4\pi)=\frac{i!}{j!(i-j)!}p_b^j(1-p_b)^{i-j}$, This result shows no atom spilling and hopping. The situation changes in the $2\pi$-winding case where the neighboring traps are separated by a finite barrier, allowing atoms to reshuffle, leading to light-induced two-body losses. Indeed, we observe large $M_{i,i-2k}$ matrix elements. We model this experiment using an estimated hopping rate from the initial temperature and calibrated one-body loss rate $p_b$ [see Fig. \hyperref[fig:matrices]{3(e)}]. The estimation of the atom hopping rate is done without any free parameters. First, the potential barrier height is extracted from the measured SO intensity pattern shown in Fig. \hyperref[fig:patterns]{1(f)}. Then the trap eigenstates are computed using the Wentzel–Kramers–Brillouin (WKB) approximation. The trap population distribution in the SO tweezers is computed considering a thermal state at loading. Finally, we suppose an equiprobable reshuffling among the traps for an atom with an energy above the potential barrier while it is held in the SO tweezers. The qualitative agreement between the model and the experiment suggests that we capture most of the mechanisms at play. The remaining disagreement might be due to the partial breakdown of the adiabatic approximation or the four-fold symmetry, and residual spilling of the atoms \cite{SM}.
	
	To explore in further detail the reshuffling mechanism, we extend the $2\pi$-winding SO tweezers duration to $80\,$ms, and reduce the potential barrier height by adiabatically lowering the ODTs laser power from $P_0$ to $P$ for 60 ms. Since the barrier height depends linearly on $P$, we expect a larger reshuffling as $P$ reduces. We confirm this general behavior by post-selecting the initial images containing only one atom. Then, by analyzing the final images, we compute the probability that the atom remains in the same trap after a round trip. This probability is lower bound by 0.25, indicating a complete reshuffling with loss of memory of the initial trap occupancy. The results are depicted in Fig. \hyperref[fig:decomp]{4(a)} for imposed separation of either 1.4$\lambda$ (red circles) or 1.7$\lambda$ (blue squares). The latter separation is characterized by a larger potential barrier of $21.4(5)\,\upmu$K, reducing the hopping and thus the reshuffling. The data are in good agreement with our adiabatic model using the initial temperature (solid areas). We note that the data presented in Fig. \hyperref[fig:decomp]{4(a)} are probabilities conditioned by the presence of an atom in the final image. The increasing single-body loss due to spilling for decreasing $P$ is shown in Fig. \hyperref[fig:decomp]{4(b)}. Roughly, the spilling becomes the dominant loss mechanism for $(P/P_0)^{-1}>4$ and it does not depend on the separation, as the trap height remains unchanged.
	
	In summary, we generated four coherent SO traps achieving super-resolution, meaning an inter-trap separation below the Sparrow limit and a sub-wavelength radial FWHM below the Abbe limit.
	In our experiment, the connectivity between traps was controlled through the relative phase of the ODTs, allowing or impeding the thermal reshuffling of atoms among the traps. Looking forward, it would be interesting to coherently control the inter-trap dynamics by bringing the atoms to the tweezers ground-state \cite{thompson2013} and study tunneling.
	The SO tweezers appear to be a promising approach that combines the best of both worlds, namely the short spacing of optical lattices and the ability to control the initial population of each trap \cite{barredo2016,endres2016}, trap geometry, and trap number \cite{nishimura2024} of optical tweezers. We believe that our results will stimulate  the investigation of larger super-resolved tweezer arrays, either using a straightforward generalization of our technique for an even number of atoms lying on a ring, or for a general geometry. The latter require advanced phase retrieval algorithms \cite{nishimura2024} or higher-dimensional structured light \cite{He2022,Ma2024}. On the experimental side, few super-resolved spots can be used for molecule assembling \cite{Liu2019}, atomtronics \cite{amico2021}, and few-body physics \cite{Murmann2015,GonzalezCuadra2023}. A ring tweezer offers also an ideal platform to verify predictions in quantum many-body scar towers in the presence of periodic boundary conditions \cite{Papic2022}.\\
	
	This work was supported by the Singapore Ministry of Education: Grant No MOE-T2EP50120-0005, and A-star NQPI: Grant No S24Q4D0002. VM and HW thank the Centre for Disruptive Photonics Technologies for financial and logistical support. The authors thank Chang Chi Kwong for careful proofreading of the manuscript.

	K.L. and V.M. contributed equally to this work, conducting the experiment and data analysis; K.L. V.M. H.W. and Y.S. designed and realized the SO holograms; all authors participated in the writing of the manuscript and D.W. supervised the project.
	
	%\bibliography{bibliography.bib}
	
	%
	
	%%%%%%%%%%%%%%%%%%%%%%%%%%%%%%%%%%%%%%%%%%%%
	%\clearpage
	%\bibliographystyle{unsrt}
\end{document}